\documentclass[12pt]{iopart}

\usepackage{graphicx}

\RequirePackage{color}

\begin{document}

\title[Leaking oscillator]{The motion of a leaking oscillator--a study for the physics class}

\author{Hilario Rodrigues, Nelson Panza, and Dirceu Portes Jr}
\address{Departmento de F\'isica \\
Centro Federal de Educa\c{c}\~ao Tecnol\'ogica do Rio de Janeiro \\
Av. Maracan\~a 229, 20271-110, Rio de Janeiro, RJ, Brazil}
\ead{harg@cefet-rj.br}

\author{Alexandre Soares}
\address{Departmento of Matem\'atica \\
Centro Federal de Educa\c{c}\~ao Tecnol\'ogica do Rio de Janeiro \\
Av. Maracan\~a 229, 20271-110, Rio de Janeiro, RJ, Brazil}

\begin{abstract}

 This work is basically about the general form of Newton's second law for variable mass problems. We develop a model for describing the motion of the one-dimensional oscillator with a variable mass within the framework of classroom physics. We present a simple numerical procedure for the solution of the equation of motion of the system to be implemented by students and teachers. Interesting qualitative concepts as well as quantitative results for the focused problem are presented.   
   The topic has pedagogical value both from theoretical and experimental point of view. However, this article considers only  theoretical aspects of the problem. The work is addressed to basic physics courses at undergraduate level.

\end{abstract}

\maketitle

\section{Introduction}\label{sec:intro}

In mechanics, variable-mass systems are systems which have mass that does not remain constant with respect to time. In such systems, Newton's second law of motion cannot directly be applied because it is valid for constant mass systems only. Instead, a body whose mass $m$ varies with time can be described by rearranging Newton's second law and adding a term to account for the momentum carried by mass entering or leaving the system \cite{Plastino,Alonso}.
     
     Due to some conceptual difficulties, this topic is not commonly  addressed in basic physics courses. So it may be pedagogically interesting to propose new approaches to the topic for students of science and engineering at the undergraduate level.

In this respect, the study of the motion of a single oscillator as its mass
varies can serve as a rich topic of discussion in a physics classroom. What happens to the equilibrium position, the amplitude, the period and frequency of a single leaking oscillator? The answer to this question is of pedagogical interest.
 
     In this article, we discuss the proper form of Newton's second law when applied to a single-degree of freedom system with a time-dependent mass. Using theoretical considerations, we present a description of the dynamics of the single one-dimensional oscillator with the mass modeled as a quadratic function of time. The obtained equation of motion is solved by using a suitable numerical procedure for given initial conditions.
 
    The work is mainly addressed to college students and teachers. The study of this topic requires acquaintance with basic concepts of calculus and physics at basic level.

\section{Newton's second law for variable-mass systems}\label{sec:sec2}

Consider a particle of mass $m$ which is moving with velocity $v$. By definition its linear momentum is $p=mv$. According to Newton's second law, the change of the linear momentum $p$ in time is determined by $F = \frac{dp}{dt}= \frac{d}{dt} (m v)$, where $F$ is the net force acting on the particle. If the particle mass is a constant (mass does not depend on time), the Newton's second law entails $F = m \frac{dv}{dt}=ma$. The last equation represents Newton's second law commonly presented in textbooks. This form is particularly useful in obtaining the equation of motion of a constant mass particle. 

    In addition, Newton's second law is invariant under the Galilean transformations, which are defined by $x^{\prime}=x-ut$, and $v^{\prime} = v-u$, where $u$ is the velocity of the primed inertial frame of reference relative to the unprimed  inertial frame. Applying the Galilean transformations to Newton's second law we get simply $F^{\prime }=F$. That is, all inertial observers must measure the same force regardless the relative speed among them.

But, what if the the particle mass is not constant in time? Can we use in this case Newton's second law forms $F = m \frac{dv}{dt}$ or even $F = \frac{d}{dt} (m v)$? The answer is no. To see that, remember that the invariance of Newton's second law enforces that the equation of motion in the primed frame of reference must retain the same form, that is:  
\begin{equation}
F^{\prime} = \frac{d}{dt} (m v^{\prime}).  \label{cmass33c}
\end{equation}
The time derivative in the right-hand side of (\ref{cmass33c}) yields
\begin{equation} 
F^{\prime} =    m \frac{dv}{dt}  + \frac{dm}{dt} \left( v - u \right),
\end{equation}
and hence 
\begin{equation}
F^{\prime} =   \frac{d}{dt}\left( mv \right)  - \frac{dm}{dt} u \ne F .
\end{equation}

\begin{figure}[htpb] 
\centering 
\includegraphics[width=.22\textheight]{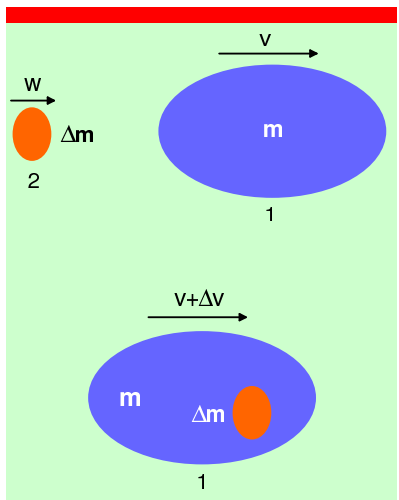}
\vspace*{-.1cm} 
\caption{(Color online) - The particle of mass $\Delta m$ and velocity $w$ collides with a particle of mass $m$ and gets stuck in it. After the process, the new particle of mass $m + \Delta m$ moves with velocity $v + \Delta v $.  \label{fig:100}}
\end{figure}

So, Newton's second law in the form $F = \frac{d}{dt} (m v)$ is not Galilean invariant when the particle mass is time dependent. In order to properly obtain the equation of motion, we have to apply the principle of conservation of linear momentum for the entire system, which is the basic principle behind the Newton's second law. Thus, consider a single-degree of freedom system with a time-varying mass $m(t)$, as illustrated in Figure \ref{fig:100}. The system (the body labeled 1 in the figure) moves with velocity $v$ at the time $t$. The particle of mass $\Delta m$ (labeled 2 in the figure) and mean velocity $w$ collides with the system during a time interval $\Delta t$,  imparting mass to the system. Assuming that the mass of the entire system is conserved during the collision, the new mass and the new velocity of the original system increase to $m + \Delta m$ and to $v + \Delta v$, respectively. The linear momentum of the system at the time $t$ is thus given by  $ p(t) = mv + \left( \Delta m \right) w$, while the new linear momentum at the time $t+\Delta t$ reads $p(t+\Delta t) = (m + \Delta m)(v +\Delta v)$. Hence, the change in the total linear momentum is
\begin{equation}
\frac{\Delta p}{\Delta t} = m \frac{\Delta v}{\Delta t} + \frac{\Delta m}{\Delta t} \Delta v - \frac{\Delta m}{\Delta t} \left(w - v\right) . \label{varp1a} 
\end{equation}

Taking the limit $\Delta t\rightarrow 0$, $\Delta m\rightarrow 0$, and $\Delta v \rightarrow 0$ in equation (\ref{varp1a}), one arrives to \cite{Plastino}
\begin{equation}
F = \frac{dp}{dt}=m\frac{dv}{dt} - \frac{dm}{dt}\left(w - v\right), \label{dpdt1a}
\end{equation}%
where $F$ is the external force acting on the system, and $w - v$ is the velocity of the incoming mass with respect to the centre of mass. Equation (\ref{dpdt1a}) can be put in the form
\begin{equation}
m\frac{dv}{dt}= F + \frac{dm}{dt} \left(w - v\right) , \label{eqmotion1a2}
\end{equation}
 Analogously, for $\frac{dm}{dt}<0$ (system losing mass) we would obtain
\begin{equation}
m\frac{dv}{dt}= F - \frac{dm}{dt} \left(w - v\right) . \label{eqmotion1a2b}
\end{equation}

Equations (\ref{eqmotion1a2}) and (\ref{eqmotion1a2b}) describe the motion of a time-varying mass particle, and represent the proper  extension of Newton's second law. The term $\frac{dm}{dt} (w-v) $ in the right-hand side should be interpreted as a real force acting on the particle, apart from the external force $F$. Also notice that equation (\ref{dpdt1a}) may be put in the form
\begin{equation}
F = \frac{d}{dt} (m v) - \frac{dm}{dt} w , \label{eqmotion1a3}
\end{equation}
which means that equation (\ref{eqmotion1a3}) recovers the form $F = \frac{d}{dt} (m v)$ in the particular case $w=0$. It is easy to prove that equation (\ref{eqmotion1a3}) is invariant under Galilean transformation.

\section{Modeling a leaking oscillator}\label{sec:sec4}

\begin{figure}[htpb]
\centering 
\includegraphics[width=.22\textheight]{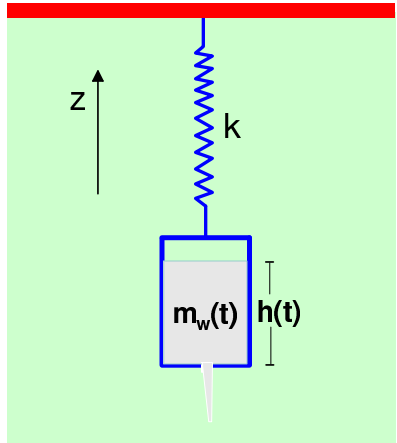}
\vspace*{-.1cm} 
\caption{(Color online) - Oscillator with a variable mass. A bucket filled with water is attached to a spring. The water flows out through a small hole in the bottom of the bucket. \label{fig:200}}
\end{figure}

In order to model the variable-mass oscillator, consider a leaking bucket of water which is attached to a spring, as illustrated in Figure {\ref{fig:200}}. The water exits out the bucket through a small hole at the bottom. Assume that the mass loss of water and the motion of the oscillator are along a line (the $z$-axis). In this situation, and ignoring friction, the system is subjected to the action of three different forces, namely, the elastic force exerted by the spring, the weight of the oscillator, and the force exerted by the leaking water. In accordance with equation (\ref{eqmotion1a2b}), the dynamical behaviour of the system is governed by the equation of motion
\begin{equation}
m \dot v = -\frac{d m}{d t} (w-v) - kz - mg, \label{eqmotion2}
\end{equation}
where $z(t)$ is the displacement of the centre of mass measured from the initial equilibrium position; $w$ is the mean velocity at which the water leaves the system; $v=\dot z$ is the velocity of the oscillator; $k$ is the stiffness coefficient of the linear restoring force; and $g$ is the acceleration of gravity.

\vspace*{.2cm}

 For the conditions $w=v=0$, $\frac{dm}{dt}=0$, and $\dot v =0$, at the time $t=0$, we obtain the initial equilibrium position 
\begin{equation}
z_{0}=-\frac{m(0) g}{k} , \label{zinit} 
\end{equation}
where $m(0)$ is the initial mass of the oscillator. If $m$ is constant, the system would oscillate around the equilibrium position $z_{0}$.

The mass of water has a quadratic dependence on the time (see appendix for details of calculation) which is given by
\begin{equation}
m_w(t) =  m_w(0) \left(1-f t\sqrt{\frac{g}{2h_{0}}}\right)^{2}, \label{masst}
\end{equation}
where $m_w(0)$ is the initial mass of water, $f=\frac{a}{A}$ is the ratio between the cross-sectional area $a$ of the hole, and the  cross-sectional area $A$ of the column of water, and $h_{0}$ is the initial height of the column of water. The mass of the oscillator is given by the summation of the mass of the bucket $m_b$, and the time-varying mass of water $m_w(t)$.  

Assuming the leaking of water occurs at a very low rate, one can neglect the effect of the first term on the right side of equation (\ref{eqmotion2}). In this approach the equation of motion reads  
\begin{equation}
 \dot v = - \frac{k}{m_b + m_w}z -  g. \label{61}
\end{equation}

 According to equation (\ref{masst}), the bucket of water is completely empty after the elapsed time given by
\begin{equation}
\tau=\frac{1}{f}\sqrt{\frac{2h_{0}}{g}} . \label{60}
\end{equation}

After the elapsed time $\tau$, the oscillations are governed by the equation of motion
\begin{equation}
 \dot v = -\frac{k}{m_b}z- g, \, \, \, \, \, \,  t\geq\tau . \label{62}
\end{equation}

\section{Numerical solution}\label{sec:sec5}

Equation (\ref{62}) represents a simple harmonic motion, whose exact solution is known and is given by a sinusoidal function. However, the exact solution of equation (\ref{61}) is beyond the scope of the present article. That does not mean we have to give up the equation  (\ref{61}). It is possible to deal with approximate numerical solutions, which can be discussed in standard physics class. With this aim, we use a simple numerical method which can be implemented, for example, in electronic calculators or even by using an Excel spreadsheet.

We start with a discrete set of equidistant instants of time $\{ t_i \}$, where $t_i=t_0 +i h$. The time step $h$ and the starting time $t_0$ are suitable chosen in order to provide results as accurate as possible. From equation (\ref{61}), we see that the time evolution of $v(t)$, the value of $\dot v$, depends on the current value $z(t)$ and the mass of water $m_w(t)$. With a sufficiently small time step $h$  the values of $v(t)$ and $z(t)$ will vary approximately linearly between two consecutive instants of time $t_i$ and $t_{i+1}$. So, we may calculate the value $v(t_{i+1})$ from  $v(t_{i})$ and the values of $z(t_i)$ and $m_w(t_i)$, setting $\dot v \approx (v_{i+1} -v_i)/h$. Analogously, we may compute the value of $z(t_{i+1})$ from $z(t_{i})$ and $v(t_i)$, by setting now $\dot z \approx (z_{i+1} -z_i)/h$. 

So, for the time interval $0\leq t\leq\tau$, we first compute $v(t)$ at the time step $t_{i+1} \doteq t_i + h$:
\begin{equation}
v_{i+1}=v_{i} - h \left(\frac{k}{m_b + m_w(t_i)} z(t_i) +  g \right) , \label{z11}
\end{equation} 
and thus one computes the value of $z(t)$
\begin{equation}
z_{i+1} = z_i + hv_{i}, \label{az11}
\end{equation}
with $z_{i+1} = z(t_{i+1})$, $v_{i+1} = v(t_{i+1})$, and $m_{i} = m_w(t_i)$, as given by equation (\ref{masst}).

Thus, the output of each loop are the values $z_{i+1}$ and $v_{i+1}$, provided the values of $v_{i}$, $z_{i}$, and $m_{i}$ of the previous loop. 

In order to compute the dynamical evolution for $t_i>\tau$, replace equation (\ref{z11}) by 
\begin{equation}
v_{i+1} = v_i - h \left( \frac{k}{m_b}z_i + g \right) , \label{62bn}
\end{equation}
according to (\ref{62}). 

The approach of numerically solving ordinary differential equations outlined above is known as Euler's method. Eventually, more sophisticated methods, as the family of Runge-Kutta methods, are extensions of these basic ideas.

\section{Numerical results}\label{sec:sec6}

  In this section we present results obtained for the case of a bucket of mass $m_b = 1.0$ kg, filled with the initial mass of water of $m_w(0) = 10$ kg with a column of initial height $h_0 = 0.5$ m. The bucket of water is attached to the spring of stiffness coefficient $k=100$ N/m. The mass of the oscillator is given at every time by $m(t) = m_b + m_w(t)$, $m_w(t)$ being the time-varying mass of water.

The algorithm for solving the equations (\ref{61}) and (\ref{62}) comprises the following steps. First, assign initial values to all variables: the starting time $t_0 = 0$; the initial position $z=z(0)$; the initial velocity $v(0)$; the initial mass of water $m_w(0)$; the initial height of the water column $h_0$; the value of the ratio between the cross-sectional areas $f=\frac{a}{A}$.   Assign values to constants $g$, $k$, and the mass of the bucket $m_b$. Now, compute the velocity $v_{i+1}$ given by equation (\ref{z11}) valid for $0\leq t \leq \tau$, or by (\ref{62bn}) for $t_i > \tau$. Then, compute the position $z_{i+1}$, given by equation (\ref{az11}). At each step the elapsed time is incremented by the time step $h$. 

\begin{figure}[htpb]
\centering 
\includegraphics[width=.60\textheight]{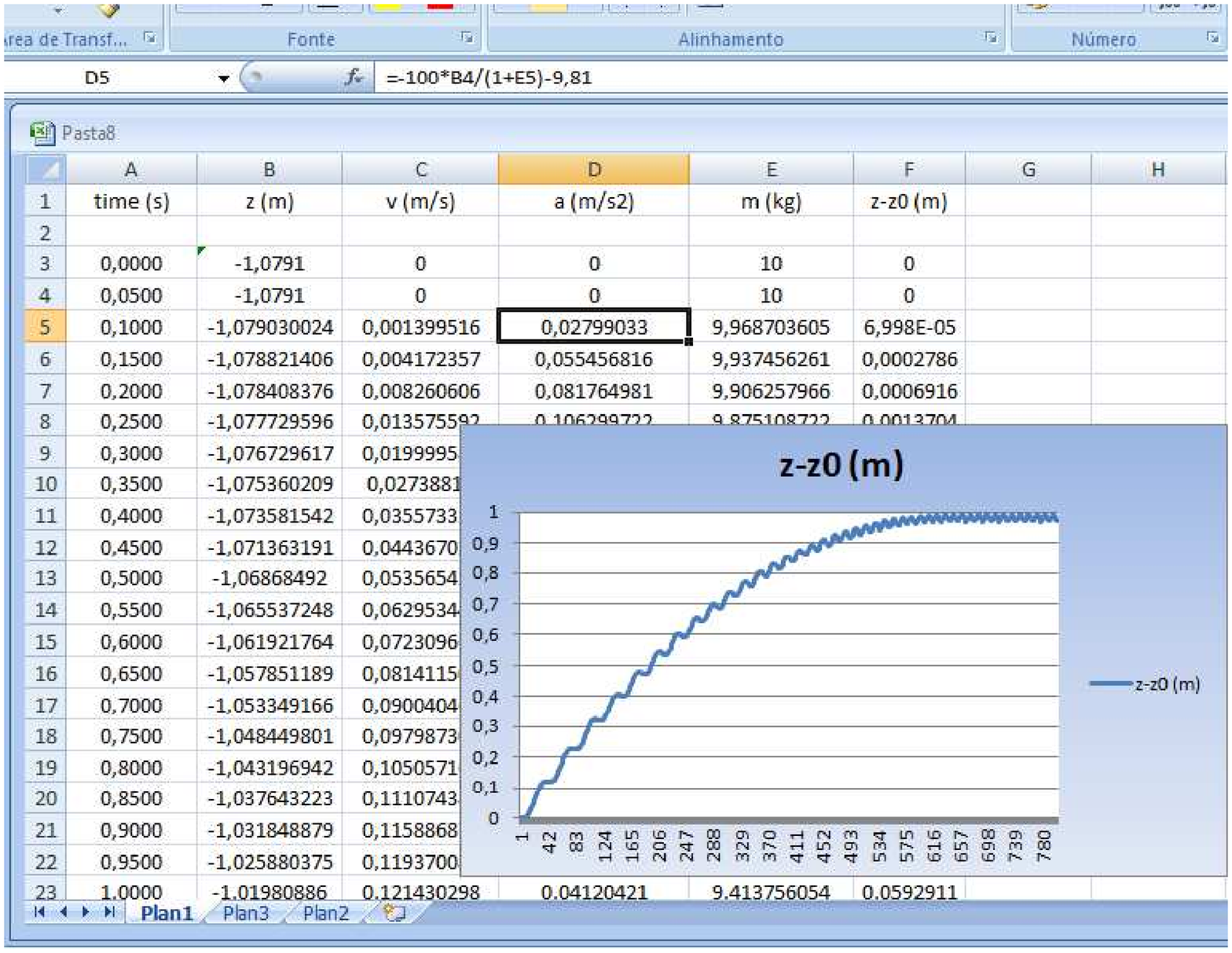}
\vspace*{-.2cm} 
\caption{(Color online) - Screen shot of the spreadsheet implementation of the numerical procedure described in Section \ref {sec:sec5} and detailed above. The columns contain the sequences $t_i$, $z_i$, $v_i$, $a_i$, $m_w(t_i)$ and the $z-z_0$, with $z_0$ given by equation (\ref{zinit}). Subsequent values of $z_{i+1}$ and $v_{i+1}$ are computed from cell entries in rows ($i-1$). The plot depicts the obtained results for $z(t)-z_0$.  The starting values of $z$, $v$, and $m_w$ are $0$ m, $0$ m$\cdot$s$^{-1}$ and $10$ kg, respectively. The used values of the other parameters are $g=9.8$ m$\cdot$s$^{-2}$, $k=100.0$ kg$\cdot$s$^{-1}$, $m_b = 1.0$ kg (mass of the bucket), and $h = 0.05$ s (the time step). \label{fig:30}}
\end{figure}

 We start with the initial condition $z(0)=0$ and $v(0)=0$. Thus, the change of the dynamical state of the system is purely caused by the change in mass of the oscillator with time. We also compute the kinetic energy, $T=mv^2/2$, the elastic potential energy, $U_k=k z^2/2$, and the gravitational potential energy, $W = m g (z-z_0)$. The mechanical energy of the system is thus given by the summation $E= T + U_k + W$. 
 
  For the value of the ratio $f=a/A=0.01$, for example, the water takes tens of seconds to exit the bucket completely. So, this number does not demand a  huge computational time, providing very accurate numerical results. In the simulations presented here, we adopt the value $h = 0.05$ s for the time step. With this value, we need nicely a few hundred loops to carry out the simulation.

The screen shot shown in Figure \ref{fig:30} illustrates the spreadsheet implementation of the numerical integration of equations (\ref{61}) and (\ref{62}) using the Euler method. We have implemented the essentially same algorithm in a Fortran compiler, and some obtained results are shown in the figures below.  

\begin{figure}[htpb]
\centering 
\includegraphics[width=.30\textheight]{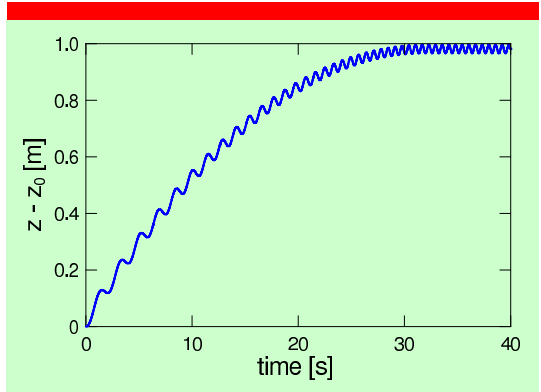}
\vspace*{-.1cm} 
\caption{(Color online) - Position as a function of time for $f=0.01$. The used values of the other parameters are $g=9.8$ m$\cdot$s$^{-2}$, $k=100.0$ kg$\cdot$s$^{-1}$, $m_b = 1.0$ kg (mass of the bucket), $m_w(0) = 10.0$ kg (initial mass of water), $h(0) = 0.5$ m (initial height of the water column), and $h = 0.05$ s (the time step).   \label{fig:40}}
\end{figure}

\begin{figure}[htpb]
\centering 
\includegraphics[width=.30\textheight]{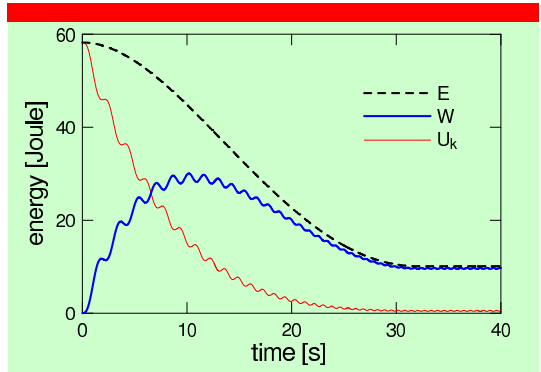}
\vspace*{-.1cm} 
\caption{(Color online) - Energy as a function of time for $f=0.01$. The values of the other parameters are the same used in the previous figure.   \label{fig:50}}
\end{figure}

Figure \ref{fig:40} depicts the position of the oscillator as a function of time for the adopted values of the model parameters outlined in the caption of the figure. Notice as the ''instantaneous'' equilibrium position of the oscillator moves upward while the water within the bucket flows out. The oscillations are obviously caused by the action of the restoring force, as the mass of the oscillator decreases. In special, one notices that the amplitude of the oscillations decreases, while the frequency increases as the mass of the oscillator decreases. The final equilibrium position, around which the system oscillates for  $t>\tau$, can be computed as  $z_{eq}=\frac{m_w(0)g}{k}$, which in the present case for $m_w(0)=10$ kg and $k=100$ N$/$m gives the value $0.98$ m. 
  
  Figure \ref{fig:50} shows the behaviour of the energy of the oscillator as a function of time for the same set of values of the parameters used in Figure \ref{fig:40}. We conclude that part of energy of the system is dissipated due to loss of mass, that is,  the leaking oscillator behaves like a damped oscillator.

\section{Conclusions}\label{sec:sec8}

In this work, we discuss the appropriate form of Newton's second law applied to single-degree of freedom systems with a time-variable mass. We present a set of equations which are used to model the dynamics of a one-dimensional oscillator with a time-varying mass. The dependence of the mass on the time is taken into account, by means of a simple modeling (the leaking bucket of water) where the mass of the oscillator has a quadratic dependence on time.

The resulting equation of motion is numerically solved in terms of the improved Euler method, and some results for chosen values of the model parameters have been  presented and discussed in the text. 

According to the results obtained by the numerical simulations, the system shows a typical  oscillatory behaviour with "amplitude" and "frequency" which vary as the water leaves the bucket. At the end, there remains only the bucket that oscillates like a one-dimensional harmonic oscillator with constant amplitude and frequency.

At this point, we point out that the quadratic dependence of mass on time is only a motivator for the leaking oscillator problem, treated here as a purely theoretical problem. Therefore, this result should be considered within its appropriate limitations.  Probably, when the bucket is moving, going up and down with the oscillations, the flow rate through the hole could be seen to change as well, deviating slightly from the results obtained here. In other words, we are ignoring the fact that the bucket, as well as the water within it, are  accelerating frames. However, we can admit that the quadratic dependence of mass must work as a reasonable approximation in the case the loss of water occurs at a very low rate, and the oscillating bucket experiences smooth motions as investigated in this article. As a suggestion, the model could be investigated experimentally using a motor and a leaking bucket of water, for example, in order to validate or not the assumptions made for the present model for the quadratic dependence of mass.
  
 This study, despite its simplicity, is intended to be used as a useful approach for students get acquainted with the physics of systems with time-varying mass at the undergraduate level.

\bigskip
\bigskip

\bigskip

\appendix {\bf Appendix}
\medskip

 Consider a cylindrical bucket of water at rest, with cross-sectional area $A$ filled with a water column of height $h$. At the bottom of the bucket, there is a small hole with a cross-sectional area $a$ ($ a \ll A $), through which water flows under its own weight when the hole is opened.

As depicted in Figure {\ref{fig:200}}, we place location 2 at the free liquid surface, and location 1 at the bottom of the bucket. Neglecting losses, which is reasonable if the hole is tiny and the storage bucket is large and wide, we can apply the Bernoulli equation:
\begin{equation}
p_{2}+\frac{1}{2}\rho Q^{2}+\rho g\left( z_{1} + h \right) = p_{1}+\frac{1}{2%
}\rho q^{2} + \rho g z_{1}, \label{bern1}
\end{equation}
where $p_1$ and $p_2$ are the pressure at the bottom of the bucket and at the free liquid surface, respectively. The upper part of the bucket is open to the atmosphere, and the water leaks the bucket freely through the hole in the bucket bottom. So, we have $p_1=p_2=p_0$, where $p_0$ is the local atmospheric pressure. $Q$ is the velocity at the free liquid surface and $q$ is the exit velocity of water;  $h$ is the height of the free liquid surface relative to the bottom; $\rho$ is the density of the liquid; and $z_1$ is the position of the bottom of the bucket in the $z$-axis. Because $A \gg a$ the term $Q$ can be neglected and set equal to zero. Thus, equation (\ref{bern1}) leads to 
\begin{equation}
q = \sqrt{2gh} . \label{qvel}
\end{equation}

 Notice that equation (\ref{qvel}) is valid even when the surface level is decreasing due to water leakage, provided that the time rate of change of $h$ and $Q$ is sufficiently small. 
 
 From the equation of continuity the rate of loss of mass is related to the mass flow trough the equation
\begin{equation}
\frac{dm}{dt}=- \rho q a \label{dmdt1}.
\end{equation}
On the other hand, the mass of water stored in the bucket at the time $t$ is given by  
\begin{equation}
m(t) = \rho A h . \label{mass1}
\end{equation}

Inserting (\ref{qvel}) into (\ref{dmdt1}) yields
\begin{equation}
\frac{dm}{dt} = - \rho A \sqrt{2gh}. \label{dmdt2}
\end{equation}

From (\ref{mass1}), we can put (\ref{dmdt2}) in the form 
\begin{equation}
\frac{dm}{dt}=-f \sqrt{\rho A} \sqrt{2gm} , \label{eqcin12433}
\end{equation}
where $f = \frac{a}{A}$. Thus, equation (\ref{eqcin12433}) leads to
\begin{equation}
\int_{m_w(0)}^{m_w} \frac{dm}{\sqrt{m}}=  -\int_0^t f \sqrt{\rho A} \sqrt{2g} dt , \label{eqcin12435}
\end{equation}
where $m_w(0) = \rho A h_0$. By carrying out both integrals in the equation (\ref{eqcin12435}), one obtains equation (\ref{masst}).

\end{document}